\documentclass[a4paper,10pt,twoside]{cpc-hepnp}
\usepackage{amsmath}
\usepackage{amssymb}
\usepackage{epsfig}
\usepackage{subfigure}
\usepackage{float}
\usepackage{verbatim} 
\usepackage{pstricks}
\usepackage{color}
\usepackage{slashed}
\usepackage{multirow}
\usepackage{pstricks}
\usepackage{color}
\usepackage{booktabs}
\usepackage{epstopdf}
\usepackage{multicol}

\usepackage{CJKutf8}
\newcommand{\zw}[1]{\begin{CJK}{UTF8}{gbsn}{#1}\end{CJK}}

\begin{document}

\fancyhead[c]{\small Prepared for Chinese Physics C}


\title{Constraining $qqtt$ operators from four-top production:\\
	a case for enhanced EFT sensitivity\thanks{Supported by the 100-talent
	project of Chinese Academy of Sciences }}

\author{%
	Cen Zhang(\zw{张岑})$^{1}$\email{cenzhang@ihep.ac.cn}%
}
\maketitle

\address{%
$^1$ Institute of High Energy Physics, Chinese Academy of Sciences, Beijing 100049, China\\
}

\begin{abstract}
	Recently, experimental collaborations have reported $\mathcal{O}(10)$
	upper limits on the signal strength of four-top production at the LHC.
	Surprisingly, we find that the constraining power of four-top
	production on the $qqtt$ type of operators is already competitive with
	the measurements of top-pair production, even though the precision
	level of the latter is more than two orders of magnitude better.  This
	is explained by the enhanced sensitivity of the four-top cross section
	to $qqtt$ operators, due to multiple insertion of operators in the
	squared amplitude, and to the large threshold energy of four-top
	production.  We point out that even though the dominant contribution
	beyond the standard model comes from the $\mathcal{O}(C^4/\Lambda^8)$
	terms, the effective field theory expansion remains valid for a wide
	range of underlying theories.  Considering the possible improvements of
	this measurement with higher integrated luminosity, we believe that
	this process will become even more crucial for probing and testing the
	standard model deviations in the top-quark sector, and will eventually
	provide valuable information about the top-quark properties, leading to
	significant improvements in precision top physics.
\end{abstract}

\begin{keyword}
	Top quark, LHC, effective field theory
\end{keyword}

\begin{pacs}
	14.65.Ha, 13.85.Hd
\end{pacs}

\begin{multicols}{2}

\section{Introduction}
As a top-quark factory with more than six million top-quark pairs produced at
Run-I and much more to expect in the future, the LHC is an ideal place to probe
the top-quark properties.  In proton-proton collision, most top quarks are
produced in $t\bar t$ pairs. Single top production has the second largest cross
section, which is about one third of the $t\bar t$.  More recently, associated
production modes such as $t\bar t+X$ and single $t+X$, where $X$ is a gauge
boson or the Higgs boson, have also been extensively studied.  These are the
main channels that are now pushing the top-quark physics into a precision era
\cite{Patrignani:2016xqp}.

Attention has also been paid to the four-top production mode, $pp\to tt\bar
t\bar t$, which, despite its tiny rate ($\approx9$ fb
\cite{Bevilacqua:2012em,Alwall:2014hca}) in the standard model (SM), i.e.~five
orders of magnitude lower than $t\bar t$ production (832 pb,
\cite{Czakon:2013goa,Czakon:2011xx}), is particularly sensitive to new physics.
It has been noticed that the total rate of this process can be enhanced
significantly in many scenarios beyond the standard model (BSM)
\cite{Lillie:2007hd,Pomarol:2008bh,Kumar:2009vs,Cacciapaglia:2011kz,Perelstein:2011ez,
AguilarSaavedra:2011ck,Beck:2015cga,Dev:2014yca,Acharya:2009gb,Gregoire:2011ka,Degrande:2010kt,Cao:2016wib}.
This can be due to the direct production of new resonant states which
subsequently decay into tops, or to the contribution from contact four-top
operators, which rises as the energy grows.  These operators are not directly
constrained by other processes at the tree level, and therefore the four-top
channel may be the first place to see their effects.

Nevertheless, a comprehensive model-independent study of four-top production in
the context of the standard model effective field theory (SMEFT) approach
\cite{Weinberg:1979sa,Leung:1984ni,Buchmuller:1985jz} has not yet appeared in
the literature.  This is not surprising.  The SMEFT framework aims to probe the
indirect effects from BSM models that are beyond the direct reach of the LHC.
These effects are expected to show up as relatively small deviations from the
SM prediction, and therefore the most powerful approach is to combine all
available precision measurements and perform global analyses.  In the top-quark
sector, such analyses are often based on the most precise ones, such as
top-pair and single-top cross sections and distributions, branching ratio
measurements, and recently also on associated production modes such as $t\bar
tZ$ and $t\bar t\gamma$, see e.g.~Refs.~\cite{Buckley:2015nca,Buckley:2015lku}
for a recent global fit.  The four-top production, on the other hand, is still
far from being precise.  The process has been searched for in a series of
experimental reports
\cite{ATLAS:2012hpa,CMS:2013xma,Aad:2015kqa,TheATLAScollaboration:2016gxs,ATLAS:2016gqb,
ATLAS:2016btu,Sirunyan:2017tep,Sirunyan:2017uyt,Aaboud:2017faq,Alvarez:2016nrz},
and the best upper limit from Ref.~\cite{Sirunyan:2017uyt} is about 4.6 times
the SM signal.  Naively, one would not expect an $\mathcal{O}(10)$ upper bound
to provide competitive information with respect to all the other precise
measurements, except for the four-top operators that are not directly probed
elsewhere.

The goal of this work is to demonstrate that this is not the case.  For a very
important class of operators, namely the contact four-fermion interactions with
two top quarks and two light quarks, $qqtt$, we will show that the four-top
process, with only a $\mathcal{O}(10)$ upper bound, is as powerful as $t\bar t$
measurements with a percentage error.  This constraining power is
due to an enhanced sensitivity of four-top production, which comes from the
fact that its cross section can depend on up to the fourth power of the
operator coefficients, which scales like $(CE^2/\Lambda^2)^4$, where $E$ is
the energy of the process, and $C/\Lambda^2$ is the coefficient of an $qqtt$
operator.  Given the large energy scale related to this process, and the
current limits on the coefficient $C/\Lambda^2$, the factor
$(CE^2/\Lambda^2)^4$ significantly enhances the sensitivity of four-top process
to the $qqtt$ operators.  We will also show that the validity of SMEFT and its
perturbativity can be guaranteed by imposing an analysis cut on the center of
mass energy of the process at a few TeV, without reducing the enhancement
factor too much, and thus the resulting constraints apply to BSM theories that
live above this energy scale, if certain assumptions are made 
to justify the omission of operators at dim-8 and higher.

For comparison, we also consider the $t\bar t$ observables at the LHC, and
study the corresponding exclusion limit on the same class of operators.  These
observables have been incorporated in a global fit by the authors of
Refs.~\cite{Buckley:2015nca,Buckley:2015lku}.  In this work, however, the
approach we follow is quite different, mainly because we are interested in the
enhancement effect of higher powers of $CE^2/\Lambda^2$.  Even in $t\bar t$
measurements, the squared term from dim-6 operators cannot be neglected with
the current precision, and therefore instead of the four linear combinations of
$qqtt$ operators used in Refs.~\cite{Buckley:2015nca,Buckley:2015lku} (defined
in Ref.~\cite{Zhang:2010dr}), we will have to include the complete set of 14
$qqtt$ operators.  A global fit, including the main cross section and asymmetry
measurements, as well as a differential cross section measurement, will be
performed to derive the global constraints in these 14 directions. These
constraints will then be compared with those from the four-top production in
the same directions.  For the latter process, we will also consider the impact
of including the full set of $tttt$ type four-fermion operators, which might be
generated together with the $qqtt$ operators, when heavy mediator particles in
the full theory are integrated out.  Note that RG-induced constraints are also
available on the $qqtt$ operators \cite{deBlas:2015aea}, but they are typically
considered as indirect constraints.

In this work, our numerical approach will be fully based on the {\sc
MadGraph5\_aMC@NLO} framework \cite{Alwall:2014hca}. We use NNPDF3.0 parton
distribution functions (PDF) \cite{Ball:2014uwa}.  A UFO model
\cite{Degrande:2011ua} that contains all 14 $qqtt$ operators and 4 $tttt$
operators is generated using the {\sc FeynRules} package
\cite{Alloul:2013bka}.  All calculations are done at the leading order (LO).
For the four-top production, we assume that a SM $K$-factor of about 1.4
\cite{Bevilacqua:2012em,Alwall:2014hca} at the next-to-leading order (NLO) can
be applied also to the operator contributions.  This might not be a good
approximation (see Ref.~\cite{Bylund:2016phk} for an example), but is the best
we can do given that the NLO prediction for all $qqtt$ operators are not yet
available.\footnote{An NLO implementation of the four-fermion top operators
	based on the {\sc MadGraph5\_aMC@NLO} framework is in progress
	\cite{prep}.}  
The corresponding theoretical error at the NLO is about $\sim30\%$, much
smaller than the experimental ones, and so they will be neglected throughout.
Similarly, for the top-pair production mode, we always rescale
the cross sections to the state-of-the-art theory predictions, except for the
asymmetries, where the SM contribution is an NLO effect,
while those from the four-fermion operators are from LO.  We therefore only use
the LO asymmetries from the dim-6 contributions.

Regarding the experimental limits on the four-top process, we will only
consider the ones on the SM four-top production signal strength.  This implies
that the SM signal shape is always assumed.  Ideally, an experimental analysis
tailored to SMEFT operators, with various cuts on the center-of-mass energy to
ensure the validity of the effective theory expansion, would be the best for
our purpose.  This however has not been done for the $qqtt$ operators.  If one
naively applies the bound on the SM cross section, the limits on BSM will be
more conservative, as in general the effective operators lead to harder energy
distributions.  This is indeed the case for the four-top operators, as have
been considered in the experimental analyses in
Refs.~\cite{Aad:2015kqa,TheATLAScollaboration:2016gxs,ATLAS:2016gqb,ATLAS:2016btu}.
Furthermore, applying the upper bound of the total cross section on the
fiducial cross section below some center-of-mass energy cut to ensure the SMEFT
validity will also make the results conservative.  Still, even these
conservative constraints on $qqtt$ operator coefficients already compete with
those from $t\bar t$ measurements, so they are sufficient for the goal of this
work. One should keep in mind that further improvements from the experimental
side are possible.

The paper is organized as follows.  In Section {\ref{sec:setup}} we present the
relevant dim-6 operators in this work.  In Section {\ref{sec:sensi}} we explain
the enhanced sensitivity of the four-top process, and discuss the validity
range of the EFT.  We compare the constraining powers of the four-top and
$t\bar t$ cross sections in Section {\ref{sec:signal}}.  Section
{\ref{sec:global}} is devoted to a global fit using $t\bar t$ measurements,
which will be compared with the fully marginalized constraints from four-top
cross section.  In Section {\ref{sec:conclusion}} we conclude.

\section{The four-fermion operators}
\label{sec:setup}

In this work we are interested in the four-fermion operators that involve two
top quarks and two light quarks.  This is an important class of operators, as
they are common in BSM models where new heavy states couple to both $t\bar t$
and $q\bar q$, or $q\bar t$ and $t\bar q$ currents. 

Assuming an U(2)$^{3(u,d,q)}$ flavor symmetry for the first two generations,  
the full set of $qqtt$ operators at dim-6 can be written as follows
\begin{eqnarray}
\mathcal{O}^{(8,3)}_{Qq}&=&\left(\bar Q_L \gamma_\mu T^a \tau^i Q_L\right) \left(\bar q_L \gamma^\mu T^a \tau^i q_L\right)
\label{eq:8a}\\
\mathcal{O}^{(8,1)}_{Qq}&=&\left(\bar Q_L \gamma_\mu T^a Q_L\right) \left(\bar q_L \gamma^\mu T^a q_L\right)\\
\mathcal{O}^{(8)}_{td}&=&\left(\bar t_R \gamma_\mu T^a  t_R\right) \left(\bar d_R \gamma^\mu T^a d_R\right)\\
\mathcal{O}^{(8)}_{tu}&=&\left(\bar t_R \gamma_\mu T^a  t_R\right) \left(\bar u_R \gamma^\mu T^a u_R\right)\\
\mathcal{O}^{(8)}_{tq}&=&\left(\bar t_R \gamma_\mu T^a  t_R\right) \left(\bar q_L \gamma^\mu T^a q_L\right)\\
\mathcal{O}^{(8)}_{Qd}&=&\left(\bar Q_L \gamma_\mu T^a Q_L \right)\left(\bar d_R \gamma^\mu T^a d_R\right)\\
\mathcal{O}^{(8)}_{Qu}&=&\left(\bar Q_L \gamma_\mu T^a Q_L \right)\left(\bar u_R \gamma^\mu T^a u_R\right)
\label{eq:8b}\\
\mathcal{O}^{(1,3)}_{Qq}&=&\left(\bar Q_L \gamma_\mu  \tau^i Q_L\right) \left(\bar q_L \gamma^\mu  \tau^i q_L\right)
\label{eq:1a}\\
\mathcal{O}^{(1,1)}_{Qq}&=&\left(\bar Q_L \gamma_\mu Q_L\right) \left(\bar q_L \gamma^\mu  q_L\right)\\
\mathcal{O}^{(1)}_{td}&=&\left(\bar t_R \gamma_\mu  t_R\right) \left(\bar d_R \gamma^\mu  d_R\right)\\
\mathcal{O}^{(1)}_{tu}&=&\left(\bar t_R \gamma_\mu  t_R\right) \left(\bar u_R \gamma^\mu  u_R\right)\\
\mathcal{O}^{(1)}_{tq}&=&\left(\bar t_R \gamma_\mu  t_R\right) \left(\bar q_L \gamma^\mu  q_L\right)\\
\mathcal{O}^{(1)}_{Qd}&=&\left(\bar Q_L \gamma_\mu Q_L \right)\left(\bar d_R \gamma^\mu  d_R\right)\\
\mathcal{O}^{(1)}_{Qu}&=&\left(\bar Q_L \gamma_\mu Q_L \right)\left(\bar u_R \gamma^\mu  u_R\right)
\label{eq:1b}
\end{eqnarray}
where $Q_L$ represents the left-handed doublet for the 3rd generation, and
$q_L$, $u_R$ and $d_R$ represent the 1st and the 2nd generation quarks.
The operators are summed over the first two generations, but we omit the
flavor indices. Other four-fermion operators are excluded by the flavor
symmetry.

For later convenience we have written the 14 operators in the form of a
top-quark vector current (color singlet or octet) contracted with a light-quark
vector current.  Their contributions to both $q\bar q\to t\bar t$ and $q\bar
q\to tt\bar t\bar t$ are independent of each other.  One could also count the 14
degrees of freedom in a more physical way:
\begin{itemize}
	\item Both the light and the heavy quark currents can be either left- or
		right-handed.  This counts 4 degrees of freedom.
	\item The light quark can be up/charm or down/strange.  This leads to
		8 in total.
	\item SU(2)$_L$ symmetry requires that $u_Lu_Lt_Rt_R$ and $d_Ld_Lt_Rt_R$ 
		interactions have the same coefficient.  This reduces the number
		to 7.
	\item With two possible color structures, i.e.~singlet and octet,
		the total number of degrees of freedom is 14.
\end{itemize}
In the $t\bar t$ process, the cross section can be written as a quadratic function
of 14 operator coefficients:
\begin{flalign}
	\sigma&=\sigma_\mathrm{SM}+\sum_i \dfrac{C_{i}}{\Lambda^2}\sigma_i
	+\sum_{i\le j} \dfrac{C_iC_j}{\Lambda^4}\sigma_{ij}
\end{flalign}
If one truncates the function and keeps only the interference term, then
the 8 color-singlet operators, Eqs.~(\ref{eq:1a})-(\ref{eq:1b}), do not give
any contribution at the LO. Furthermore, without information from the decay of the tops,
the LLLL (LLRR) interactions cannot be distinguished from the RRRR (RRLL) operators.
Therefore only 4 degrees of freedom can be observed \cite{Zhang:2010dr}, which
significantly simplifies the analysis.  However, the current limits on the
operator coefficients $C/\Lambda^2$ indicate that the dim-6 squared terms
are not negligible, and so the full set of 14 operator need to be included in
$t\bar t$ production.  The four-top production mode is similar, and in particular,
there the dominant terms may come from the fourth power of dim-6 coefficients.

Fortunately, as we will see in Section {\ref{sec:signal}}, the SMEFT
analysis with all 14 operators can be simplified by observing that these
operators can be divided into three categories according to the flavor of the
light quarks, without any interference effect across: 
\begin{enumerate}
	\item $u_R$:
	\begin{equation}
		\mathcal{O}^{(8)}_{tu}, \mathcal{O}^{(1)}_{tu}, \mathcal{O}^{(8)}_{Qu}, \mathcal{O}^{(1)}_{Qu};
		\label{eq:c1}
	\end{equation}
	\item $d_R$:
	\begin{equation}
	\mathcal{O}^{(8)}_{td}, \mathcal{O}^{(1)}_{td}, \mathcal{O}^{(8)}_{Qd}, \mathcal{O}^{(1)}_{Qd};
		\label{eq:c2}
	\end{equation}
	\item $q_L$:
	\begin{equation}
	\mathcal{O}^{(8,3)}_{Qq}, \mathcal{O}^{(8,1)}_{Qq}, \mathcal{O}^{(1,3)}_{Qq}, \mathcal{O}^{(1,1)}_{Qq},
	\mathcal{O}^{(8)}_{tq}, \mathcal{O}^{(1)}_{tq}.
		\label{eq:c3}
	\end{equation}
\end{enumerate}
Furthermore, the operators in the first two categories can be easily related to
those in the last category by parity.  This implies that one analysis with 14
operators can be simplified into two independent ones, each with only 4
operators, from the 1st or the 2nd category, and parity can be used
to derive results for the 3rd category.  This is one of the reasons for
choosing the operator basis given by Eqs.~(\ref{eq:8a})-(\ref{eq:1b}).  As we
will see, this simplification is very important for analyzing the four-top
production process, as there the cross section is a quartic function of 14
operators, with a large number of interference terms.

We also consider the operators that consist of four top quarks. These four-top
operators are important because unlike the $qqtt$ ones, they are bound to be
generated as long as there are BSM particles coupled to the top quark.  The
four-top production is the first process to directly probe them (see, for
example, discussions in
Refs.~\cite{Pomarol:2008bh,Kumar:2009vs,Degrande:2010kt,Servant:2010zza}).  In
this work we will also provide constraints on these operators.  Note that our
main goal is to derive constraints on the $qqtt$ operators, however, reliable
constraints need to be obtained by marginalizing over other operators that
enter the same process.  This is the main reason to study the contribution from
four-top operators, as we want our conclusion to be independent of their sizes.

Five such operators exist in the so-called Warsaw basis~\cite{Grzadkowski:2010es}:
\begin{equation}
	\mathcal{O}^{(1)(3333)}_{qq}, \mathcal{O}^{(3)(3333)}_{qq},
	\mathcal{O}^{(3333)}_{uu}, \mathcal{O}^{(1)(3333)}_{qu},
	\mathcal{O}^{(8)(3333)}_{qu}\,.
\end{equation}
Among them only four are independent in four-top production, which we define as
\begin{flalign}
	&\mathcal{O}^{(+)}_{QQ}\equiv
	\frac{1}{2}\mathcal{O}^{(1)(3333)}_{qq}+\frac{1}{2}\mathcal{O}^{(3)(3333)}_{qq}\,,
	\label{eq:4t1}\\
	&\mathcal{O}_{tt}\equiv \mathcal{O}_{uu}^{(3333)}\,,
	\\
	&\mathcal{O}_{Qt}^{(1)}\equiv \mathcal{O}_{qu}^{(1)(3333)}\,,
	\\
	&\mathcal{O}_{Qt}^{(8)}\equiv \mathcal{O}_{qu}^{(8)(3333)}\,,
	\label{eq:4t2}
\end{flalign}
while the remaining degree of freedom is chosen as
\begin{equation}
	\mathcal{O}^{(-)}_{QQ}\equiv
	\frac{1}{2}\mathcal{O}^{(1)(3333)}_{qq}-\frac{1}{2}\mathcal{O}^{(3)(3333)}_{qq}\,,
\end{equation}
with no contribution to the process.  

These four-top operators in general interfere with the other $qqtt$ operators
in the four-top production.  For a complete analysis, one will have to consider
each category in Eqs.~(\ref{eq:c1})-(\ref{eq:c3}) together with these four
operators.  Parity relation still holds, under which $C^{(+)}_{QQ}$ and
$C_{tt}$ is exchanged (neglecting contributions initiated by two $b$ quarks).

The relation between our four-fermion operator basis and the more standard
basis, i.e.~the Warsaw basis in Ref.~\cite{Grzadkowski:2010es}, is given in
Appendix A.

Finally, we briefly explain the notation used in this work. The coefficients of
dim-6 operators are denoted as $C/\Lambda^2$.  One should keep in mind that the
$C$ and $\Lambda$ individually do not have any physical meaning.  Only their
combination is a physical quantity.  We define
\begin{equation}
	\tilde C_i\equiv\frac{C_i(1\ \mathrm{TeV})^2}{\Lambda^2}\,
\end{equation}
so that constraints on SM deviations can be conveniently quoted in terms of
$\tilde C$.  The values of $\tilde C$ are constrained by experiments and are
model-independent.  On the other hand, we use $\Lambda_{NP}$ to denote the
characteristic scale at which the new physics resides.  This is not a
model-independent quantity, but it is useful for defining the range of validity
of the EFT expansion, which requires $E<\Lambda_{NP}$, where $E$ is the typical
energy transfer in the process of interest.

\section{Sensitivity and EFT validity}\label{sec:sensi}

To briefly explain the sensitivity of the four-top process to four-fermion
$qqtt$ operators, let us take $\mathcal{O}^{(8)}_{tu}$ as an example.  This
operator represents a contact interaction between a color octet right-handed
up-quark current and a color octet right-handed top-quark current.  We first
consider the $t\bar t$ process.  $t\bar t$ measurements so far impose the
tightest bounds on $qqtt$ operators.  The LO cross section at 8 TeV, rescaled
to the next-to-next-to-leading order (NNLO) prediction including the
resummation of next-to-next-to-leading logarithmic (NNLL) soft gluon terms
\cite{Czakon:2013goa,Czakon:2011xx}, is numerically given by (in pb):
\begin{flalign}
	252.9+2.94\tilde C^{(8)}_{tu}+0.411 {\tilde C}^{(8)}_{tu}{ }^2
	\,.
	\label{eq:ttxsec}
\end{flalign}
Using the combined ATLAS and CMS measurement on $t\bar t$ inclusive cross
section \cite{CMS:2014gta}, we find the following bounds
\begin{flalign}
	-11.8<{\tilde C}^{(8)}_{tu}<4.6
\end{flalign}
at the 95\% confidence level (CL).  Interestingly, both the upper and the lower
limits on ${\tilde C}^{(8)}_{tu}$ come only from the upper bound of the cross
section.  In particular for the lower limit ${\tilde C}^{(8)}_{tu}=-11.8$, the
squared term in Eq.~(\ref{eq:ttxsec}) already dominates over the
interference.

As we have mentioned in the introduction, it is this same effect, i.e.~the
dominance of terms with higher powers in ${\tilde C}$, that enhances the EFT
sensitivity of four-top production.  In particular, at the LO, the 14 $qqtt$
type operators can be inserted at most twice in the amplitude. The squared
amplitude at LO is thus a quartic function with 14 arguments:
\begin{flalign}
	O&=O_\mathrm{SM}+\sum_i \dfrac{C_{i}}{\Lambda^2}O_i
	+\sum_{i\le j} \dfrac{C_iC_j}{\Lambda^4}O_{ij}
	+\sum_{i\le j\le k} \dfrac{C_iC_jC_k}{\Lambda^6}O_{ijk}
	\nonumber\\&
	+\sum_{i\le j\le k\le l} \dfrac{C_iC_jC_kC_l}{\Lambda^8}O_{ijkl}\,,
\end{flalign}
where $O$ represents any observable.  Focusing again on $\mathcal{O}^{(8)}_{tu}$,
without worrying about EFT validity for the moment, the LO total cross section is
(in fb)
\begin{equation}
	6.1+0.10{\tilde C}^{(8)}_{tu}+0.081{\tilde C}^{(8)}_{tu}{}^2
	+0.016{\tilde C}^{(8)}_{tu}{}^3+0.0048{\tilde C}^{(8)}_{tu}{}^4
	\,.
	\label{eq:ttttxsec}
\end{equation}
The CMS search presented in Ref.~\cite{Sirunyan:2017uyt} gives an upper bound
on the signal strength of four-top process, $\mu<4.6$.  Naively applying this 
result to Eq.~(\ref{eq:ttttxsec}), we find the following constraints
\begin{flalign}
	-8.8<{\tilde C}^{(8)}_{tu}<7.1\,,
\end{flalign}
which are already complementary to the previous constraints from $t\bar t$.  Note,
however, that when these constraints are saturated, it is the ${\tilde
C}^{(8)}_{tu}{}^4$ term that gives the dominant contribution.  Had we
truncated Eq.~(\ref{eq:ttttxsec}) to, say, the linear term in ${\tilde
C}^{(8)}_{tu}$, the resulting constraints would have been more than one order of
magnitude worse.  This implies that the four-top process has an enhanced
sensitivity to $qqtt$ operators, due to the contribution from higher power
terms in $\tilde C$, and this is why such a process with only a
$\mathcal{O}(10)$ upper bound on its signal strength can beat the $t\bar t$
measurement with a precision at the percentage level.  Note that which term
dominates depends on the size of the $\tilde C^{(8)}_{tu}$, and is therefore
related to the current experimental bounds.  The quartic term
dominates if $|\tilde C^{(8)}_{tu}|>4.1$, while the quadratic one dominates if
$1.2<|\tilde C^{(8)}_{tu}|<4.1$.  As the experimental constraints continue to
improve in the future, the situation might change.  Also note that a similar
effect, i.e.~the dominance of the quadratic term, has been observed in multijet
production \cite{Krauss:2016ely}.

\begin{center}
\includegraphics[width=.8\linewidth]{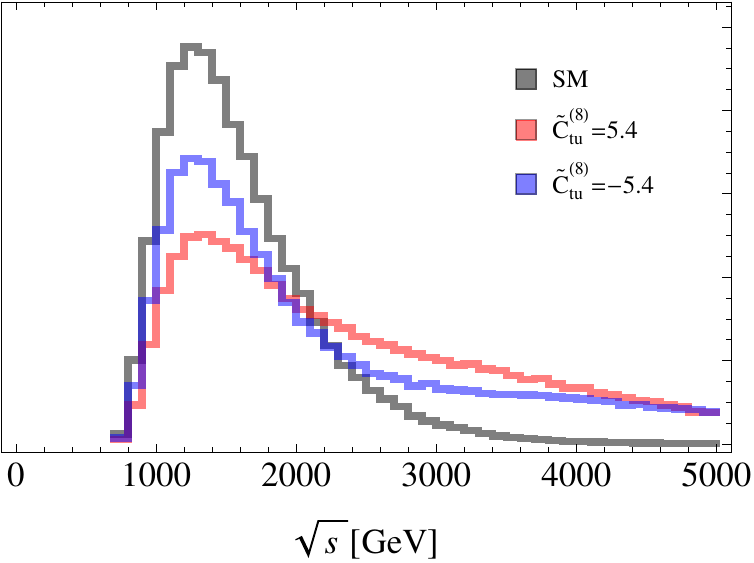}
\figcaption{\label{fig:minv} Center-of-mass energy distribution of four-top
production, normalized, to illustrate the typical energy scale of this process.
Results are shown for the SM case and for $\tilde C_{tu}=\pm5.4$ respectively.}
\end{center}

The above observation however leads to two questions: why the high power terms
dominate, and whether the SMEFT expansion is still valid.  The first question is
mostly explained by the large energy scale related to the four-top
process.  The threshold of four-top production is $4m_t\approx690$ GeV.  Most
signal events have a typical center of mass energy of $\gtrsim
\mathcal{O}$(1) TeV, depending on the value of the operator coefficients,
as illustrated in Figure~\ref{fig:minv}. The series in Eq.~(\ref{eq:ttttxsec})
comes from multiple insertion of the four-fermion effective interaction in the
squared amplitude, and by power counting each insertion corresponds to a factor
of $CE^2/\Lambda^2$, where $E$ is the characteristic energy of the process.  The
current constraints on $C/\Lambda^2$ then implies that
\begin{flalign}
	\frac{CE^2}{\Lambda^2}> 1
	\label{eq:efactor}
\end{flalign}
and so terms with the highest power in $C/\Lambda^2$ are supposed to 
dominate.\footnote{Eq.~(\ref{eq:efactor}) with $E\approx\sqrt{s}$ tends to
overestimate the effective contribution.  The reason is that the energy transfer 
at the effective vertices is often less than $\sqrt{s}$.  The only
configuration where the energy transfer is equal to $\sqrt{s}$ is the case
where the two initial quarks enter the same effective vertex, which then produces
$t^*\bar t\to t\bar tt\bar t$, but in this case the squared amplitude
can depend on at most two powers of $Cs/\Lambda^2$.
Still, in this process either $(\frac{CE^2}{\Lambda^2})^4$ with
$E\lesssim\sqrt{s}$ or
$(\frac{Cs}{\Lambda^2})^2$ represents a large factor.}
Note that this is not true for all operators.  For example, another
important operator that enters both $t\bar t$ and four-top production channels is
the top-quark chromo-magnetic dipole operator,
\begin{equation}
	\mathcal{O}_{tG}=y_tg_s(\bar Q\sigma^{\mu\nu}T^At)\tilde\phi
	G_{\mu\nu}^A\,.
\end{equation}
The contribution of this operator does not scale as $CE^2/\Lambda^2$ because
of the Higgs vev.  It is also better constrained by $t\bar t$ in the $gg$
initiated channel.  As a result, the four-top limit on $C_{tG}$ cannot compete
with the one from the $t\bar t$ measurement, and so we will not consider it in
this work.

The second question is more crucial.  The fact that higher power terms in
Eq.~(\ref{eq:ttttxsec}) dominate seems to imply the breakdown of the EFT
expansion, as one could ask whether the contributions from dim-8 and higher
operators can be safely ignored in an EFT expansion, given that they scale the
same way in $1/\Lambda$ as the higher-power terms in Eq.~(\ref{eq:ttttxsec}).
Therefore the validity of the EFT expansion itself needs to be justified.  Here
to make things clear, it is important to distinguish between two kinds of
``expansions''.  The EFT expansion comes from integrating out heavy degrees of
freedom at the energy scale $\Lambda_{NP}$ (to be distinguished from the
non-physical $\Lambda$), a procedure whose legitimacy is related to
$E/\Lambda_{NP}<1$.  This means that for a given process one could always
truncate the SMEFT Lagrangian at a certain dimension \cite{Appelquist:1974tg}.
This is however different than the ``expansion'' in Eq.~(\ref{eq:ttttxsec}),
which instead comes from multiple insertion of dimension-six effective
interaction and squaring the amplitude.  In this case the ``expansion
parameter'' is $CE^2/\Lambda^2>1$.  However, this second ``expansion'' is not
related to EFT validity, and is strictly speaking not even an expansion: there
are no more terms after the fourth power of $CE^2/\Lambda^2$ (at LO, with
on-shell tops and no further radiations), so there is no need to truncate.
Simply put, when $CE^2/\Lambda^2>1$ is allowed by experimental constraints, one
should only truncate the expansion in the dimension of operators, but keep all
terms in a series of $CE^2/\Lambda^2$.  The relative theory error due to
neglecting higher order terms is then controlled by $E^2/\Lambda_{NP}^2<1$,
instead of $CE^2/\Lambda^2>1$.

As a simple example of the above argument, it has been discussed in
Refs.~\cite{Biekoetter:2014jwa,Contino:2016jqw} that in a wide class of BSM
models with strongy couplings, the contribution from dim-8 operators is
subleading with respect to dim-6 squared terms, without invalidating the EFT
expansion.  An explicit example has been given in Ref.~\cite{Contino:2016jqw},
where a $2\to2$ scattering process is considered.  The SM contribution is of
order $g_{SM}^2$, while a dimension-six operator coming from integrating out
the heavy mediator can be as large as $g_*^2E^2/\Lambda_{NP}^2$, where $g_*$ is
the BSM coupling of the mediator to the SM particles. We have 
\begin{equation}
	\frac{C}{\Lambda^2}\sim \frac{g_*^2}{\Lambda_{NP}^2}\,.
\end{equation}
If the coupling $g_*$ is much larger than the SM coupling $g_{SM}$ (which is
often the case when experimental constraints are saturated, if $\Lambda_{NP}$
is kept larger than $E$), the BSM contribution dominates the SM contribution
when $(g_*/g_{SM})^2E^2/\Lambda_{NP}^2>1$, and similarly the BSM squared term
dominates over the interference term between SM and BSM.  The EFT expansion is
still valid if $E/\Lambda_{NP}<1$, because a $2\to2$ scattering at the tree
level can be enhanced at most by $g_*^2$, and therefore no $g_*/g_{SM}$ factor
exists between dimension-six and dimension-eight operators.  In general the
validity of the EFT expansion is not spoiled by a large $g_*$, or a large
$C/\Lambda^2$, because the maximum power of $g_*$ in a given process is fixed.
As a physics case, in reality the LHC sensitivities to the triple-gauge-boson
couplings are completely dominated by dim-6 squared contributions, while the
global EFT analyses can be performed without including dim-8 operators
\cite{Butter:2016cvz,Falkowski:2016cxu}.

In four-top production the situation is similar. The $q\bar q\to tt\bar t\bar
t$ amplitude can be enhanced at most by $g_*^4E^4/\Lambda_{NP}^4\sim
(CE^2/\Lambda^2)^2$, and the cross section by $(CE^2/\Lambda^2)^4$, as shown in
for example Figure~\ref{fig:fdiagram} (a). Upon integrating out the heavy
mediators, the amplitude in the EFT is described by Figure~\ref{fig:fdiagram} (b),
i.e.~with two insertions of dimension-six operators.  However, the crucial
difference here is that truncating the higher-dimensional operators is not
guaranteed as in a $2\to2$ process.  At this point, model dependent assumptions
on the underlying theories are indispensable for further discussion.  For
simplicity, and following Ref.~\cite{Contino:2016jqw}, let us assume that the
underlying theory is characterized by one scale $\Lambda_{NP}$ and one coupling
$g_*$, and that the power counting in the EFT is given by
\cite{Pomarol:2014dya}
\begin{equation}
	\mathcal{L}_{\mathrm{EFT}}=\frac{\Lambda_{NP}^4}{g_*^2}
	\mathcal{L}\left( \frac{D_\mu}{\Lambda_{NP}},\frac{g_*H}{\Lambda_{NP}},
	\frac{g_*f_{L,R}}{\Lambda_{NP}^{3/2}},\frac{gF_{\mu\nu}}{\Lambda_{NP}^2}\right)\,.
	\label{eq:eftexp}
\end{equation}
Higher dimensional operators can be constructed in different ways.  One can use
the first expansion parameter in Eq.~(\ref{eq:eftexp}), $D_\mu/\Lambda_{NP}$,
to increase the dimension without changing the field content.  This is like
expanding a heavy mediator propagator $(p^2-M^2)^{-1}=-M^{-2}\left(
1+p^2/M^2+p^4/M^4+\dots \right)$, where $M\approx\Lambda_{NP}$, so the
expansion parameter is simply $E^2/\Lambda_{NP}^2$.  In this case neglecting
higher-dimensional operators is justified.  Alternatively, one can also use
$g_*f_{L,R}/\Lambda_{NP}^{3/2}$ or $g_*H/\Lambda_{NP}$ to increase the
dimension, and the expansion parameter is enhanced by $g_*$, so
higher-dimensional operators have a chance to contribute more.  This however
cannot be done repetitively, because at some point the operator will contain
more than six fields and become irrelevant (assuming LO amplitude dominates,
and neglecting the vev as we are interested in the high-energy regime).  The
question is where to stop this $g_*$ enhanced expansion.  Note that the dim-6
contribution is dominated by amplitudes like Figure~\ref{fig:fdiagram} (b)
which already scale like $g_*^4E^4/\Lambda_{NP}^4$.  The first relevant
operator that is enhanced by $g_*^4$ is a dim-10 operator, $g_*^4f^6D$, whose
contribution scales like $g_*^4E^6/\Lambda_{NP}^6$. This is still subdominant.
For illustration we give an example in Figure~\ref{fig:fdiagram} (c) and (d),
in a model with a heavy mediator with coupling strength $g_*$.  Note that the
two-to-four process can be enhanced at most by $g_*^4$ at the tree level, and a
SMEFT operator that contains six fermions is at least at dim-10, because
odd-dimensional operators do not exist in the SM if B and L number violating
operators are ignored \cite{Degrande:2012wf,Kobach:2016ami}.  On the other
hand, dim-8 operators are enhanced at most by 3 powers of $g_*$.  Since both
the dim-8 and dim-10 contributions are less than $g_*^4E^4/\Lambda_{NP}^4$, and
further enhancement with $g_*$ beyond dim-10 is not possible without adding
more particles, we conclude that, under the above assumption, truncating the
SMEFT at dim-6 is justified.  

\end{multicols}
\bigskip

\ruleup
\begin{center}
\includegraphics[width=.8\linewidth]{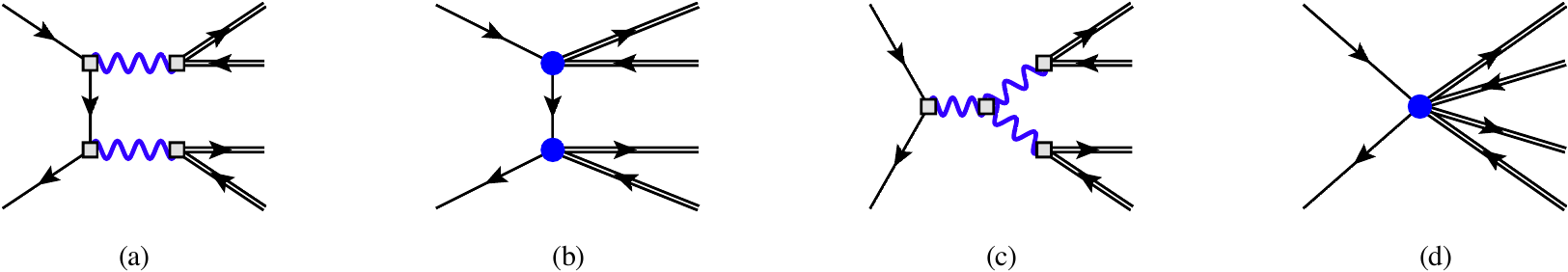}
\figcaption{\label{fig:fdiagram}
The $q\bar q\to tt\bar t\bar t$ amplitudes that are enhanced by four powers of
BSM coupling $g_*$.  Blue lines are heavy mediators. Double lines represent the
top quarks.  The square represents a $g_*$ coupling, and the blue blob
represents effective operators, coming from integrating out the mediators.
Diagrams (a), (c) describe the amplitudes in the underlying theory, while in
the EFT they respectively correspond to (b) and (d).  Diagrams (a) and (b)
correspond to two insertions of dim-6 operators.  They scale like
$g_*^4E^4/\Lambda_{NP}^4$.  Diagrams (c) and (d) correspond to one insertion of
a dim-10 operator.  They scale like $g_*^4E^6/\Lambda_{NP}^6$.
}
\end{center}
\ruledown

\bigskip

\begin{multicols}{2}

It is important to keep in mind that this conclusion is a model-dependent one.
In practice one could come up with theories with more than one scales or
couplings, where the dim-8/10 contributions might be important.  As a general
rule, when interpreting results obtained with a dim-6 SMEFT in specific models,
one always needs to check the validity of the EFT by estimating the impact of
higher dimensional contributions.

It remains to show that the validity condition, $E/\Lambda_{NP}<1$, can be
taken under control.  As proposed in Ref.~\cite{Contino:2016jqw}, the standard
way to deal with this in a hadron collider is to apply a mass cut $M_{cut}$ on
the center of mass energy of the event, or some other observable that
characterizes the energy scale of the process.  Results of the analysis should
be provided as functions of $M_{cut}$. The SMEFT approach is then valid if
results are interpreted with BSM models that satisfy $\Lambda_{NP}>M_{cut}$,
and theory errors due to missing higher dimensional terms can be estimated by
$M_{cut}^2/\Lambda_{NP}^2$.  As we have mentioned in the introduction, since
the experimental search is not carried out with this kind of strategy, we will
simply apply various $M_{cut}$ of order a few TeV on the center-of-mass energy
in our cross section calculation. The resulting fiducial cross sections are
required to be less than the upper bound set on the SM total cross section,
which then gives conservative constraints.

\begin{center}
\includegraphics[width=.83\linewidth]{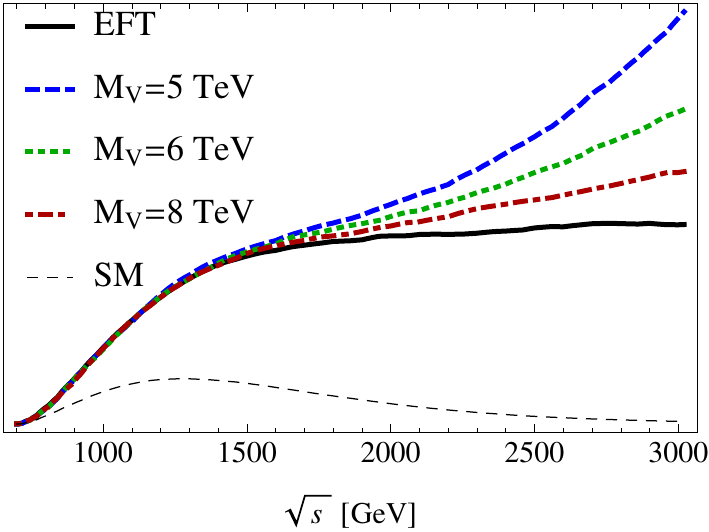}
\figcaption{\label{fig:val} Center-of-mass energy distribution of four-top
production, to illustrate the validity of EFT.
Results are shown for the EFT, the SM, and for $M_{V}=5$, 6, 8 TeV respectively.}
\end{center}

To illustrate how well the SMEFT could reproduce the full theory prediction
below $M_{cut}$, we have considered an explicit model with a heavy vector
mediator particle of mass $M_{V}$ and width $M_{V}/(8\pi)$ that couples
to the right-handed quark currents.  The coupling corresponds to $\tilde
C^{(1)}_{tu}=-4$ and $\tilde C_{tt}=-2$.\footnote{When $\Lambda_{NP}$
is large, it may seem that these values would require a large coupling strength
which is not compatible with our assumption on the width.  However it is always
possible to obtain large values of $\tilde C$ without using a very strong
coupling by arranging more than one particles, or using group factors from a
large representation, etc.}  The invariant mass distributions for $M_{V}=5$, 6, 8
TeV and for the EFT case (equivalent to $M_{V}\to\infty$) are compared in
Figure~\ref{fig:val}.  Note that the SM contribution is very small compared
with the EFT, as the latter is dominated by the dim-6 quartic contributions.
Dim-8 and higher-dimensional contributions are however subleading as
illustrated by the differences between the EFT curve and the explicit models.
For the 5 TeV case, the cross section below $M_{cut}=3$ TeV is reproduced by
the EFT with about $30\%$ error, roughly corresponds to
$(M_{cut}/\Lambda_{NP})^2$ as expected.  For larger mediator masses the EFT
approximation becomes better, which implies that the EFT validity issue will
become less severe as the measurements continue to improve in the future.  This
example corresponds to $|CE^2/\Lambda^2|\approx 36$.

The discussion in this section is based on LO accuracy.  The $qq\to
tt\bar t\bar t$ amplitude can go beyond the fourth power of $CE^2/\Lambda^2$,
if loop corrections are important.  In this case dim-8 operators are also
needed for a consistent theory prediction, as in general two dim-6 operators
can mix with a dim-8 one.  Given the precision level of the process, we simply
impose the following perturbativity condition
\begin{equation}
	\dfrac{CE^2}{(4\pi)^{2}\Lambda^{2}}<\dfrac{CM_{cut}^2}{(4\pi)^{2}\Lambda^{2}}<1\,.
	\label{eq:pert}
\end{equation}
to make sure the loop corrections due to additional insertion of effective operators
are not important.  This will be checked for typical values of $M_{cut}$.
A related issue is that jets in the final state may allow additional powers of
$CE^2/\Lambda^2$.  As an example, we find that the $tt\bar t\bar tjj$ cross section
depends on $\tilde C^{(8)}_{tu}{}^6$, with a coefficient of $5.6\times 10^{-5}$~fb.
This term is less important than the $\tilde C^{(8)}_{tu}{}^4$ term if 
$\tilde C^{(8)}_{tu}<9$, consistent with the perturbativity condition for
$M_{cut}\approx 4$ TeV.  Finally, additional powers of $CE^2/\Lambda^2$ may
come from non-top operators, if on-shell top quarks are not strictly required.
We will simply assume that these operators are more likely to be constrained by
other non-top measurements.

\section{The signal process}\label{sec:signal}

The cross section of four-top production is a quartic function of the 14
$qqtt$ operator coefficients.  Such a function in general has $C_{14+4}^4=3060$
terms.  Numerically determining this function will then require at least 3060
independent simulations at different parameter space points, which is huge
amount of work.  Fortunately,
as we have explained in Section {\ref{sec:setup}}, the procedure can be
simplified into two steps.  The first step is to determine the cross section as
functions of operators in the first two categories, separately, which requires
a minimum of only $C_{4+4}^4+C_{4+4}^4-1=139$ independent simulations.  The
second is to derive the cross section as a function of operators in the third
category, with the help of parity. Namely, if one imposes
\begin{flalign}
	C^{(8)}_{Qu}=C^{(8)}_{Qd}\,,\quad
	C^{(1)}_{Qu}=C^{(1)}_{Qd}\,,
\end{flalign} 
then the cross section is invariant under the following transformation
\begin{flalign}
	C^{(a)}_{Qu}=C^{(a)}_{Qd}\quad&\Leftrightarrow\quad C^{(a)}_{tq}\,,
	\label{eq:trans1}\\
	C^{(a)}_{\substack{tu\\td}}\quad&\Rightarrow\quad C^{(a,1)}_{Qq}\pm C^{(a,3)}_{Qq}\,,
	\\
	C^{\left(a,\substack{1\\3}\right)}_{Qq}\quad&\Rightarrow\quad 
	\frac{1}{2}\left(C^{(a)}_{tu}\pm C^{(a)}_{td}\right)\,,
	\label{eq:trans2}
\end{flalign}
where $a=1,8$.  Using these relations, the dependence on the third category
operators can be derived from that of the first two.  

When the four-top operators are included, in each category one has to consider
together the 4 $qqtt$ operators and the 4 $tttt$ operators.  The $tttt$ operators
can be inserted only once in the amplitude, if the $qqtt$ operators are not
inserted twice in the same amplitude.  This increases the total number of
independent terms to 705, which is still manageable.  The parity relations
can still be used to derive the dependence on the third category operators,
provided that
\begin{equation}
	C^{(+)}_{QQ}\Leftrightarrow C_{tt}
\end{equation}
is added to Eqs.~(\ref{eq:trans1})-(\ref{eq:trans2}).

Following the procedures described above, to determine the dependence of the
four-top cross section on the 14 $qqtt$ and 4 $tttt$ operator coefficients, we
have randomly generated
$\sim\mathcal{O}$(1000) points in the parameter space, and computed the cross
section at these points, applying $M_{cut}=2$, 3, 4 TeV respectively.  These
points are uniformly distributed roughly within the experimentally allowed
region of the coefficients. Results are then fitted to the polynomial described
above.  We have checked that the prediction of the fitted function at all these
sampled points agree with the simulation within $3\%$ error.

With this function we are ready to evaluate the constraining power of the 
signal process, and compare the constraints from four-top production with those
obtained from $t\bar t$ measurements.  For this purpose we first consider
single measurements on the $t\bar t$ total cross sections at the LHC, including:
\begin{itemize}
	\item 8 TeV ATLAS, $\sigma=242.9\pm8.8$ pb \cite{Aad:2014kva},
	\item 13 TeV CMS, $\sigma=888^{+33}_{-34}$ pb \cite{Sirunyan:2017uhy}.
\end{itemize}
Corresponding theoretical predictions at NNLO+NNLL are taken from
\cite{Czakon:2013goa,Czakon:2011xx}.  For the four-top production, we consider
the current upper bound with signal strength $\mu<4.6$ \cite{Sirunyan:2017uyt},
applying the $M_{cut}=3$ TeV cut on the center-of-mass energy.  We further
consider the projection for an integrated luminosity at 300 fb$^{-1}$,
$\mu<1.87$, estimated by Ref.~\cite{Alvarez:2016nrz}, and apply $M_{cut}=2$ TeV
and 3 TeV respectively.

In Figure~\ref{fig:limitsup} we show the resulting constraints at 95\%
confidence level, for the operators in the first category (i.e.~those that
couple to $u_R$), with two operators turned on at a time.  Results for the
other two categories are similar and are given in Appendix B.  From these
plots, our observations are the following: 
\begin{itemize}
	\item Current constraints from four-top production already provide
		competitive constraints (black dashed), which are close to, and
		in some cases better than, the constraints from the 13 TeV
		$t\bar t$ measurement with only a $4\%$ error (green dashed).
	\item The 8 TeV $t\bar t$ measurement so far gives better constraints
		(green shaded), but even these will be superseded in the future
		by an improved search/measurement of four-top production at 300
		fb$^{-1}$ luminosity with a projected $\mu<1.87$ upper bound
		(black solid), assuming $M_{cut}=3$ TeV.
	\item Lowering $M_{cut}$ to 2 TeV will give somewhat looser constraints
		(blue solid), but results can be applied to more underlying
		models where the BSM scales are not so heavy.  On the other
		hand, increasing this cut can further improve the constraints.
\end{itemize}

\begin{center}
	\includegraphics[width=\linewidth]{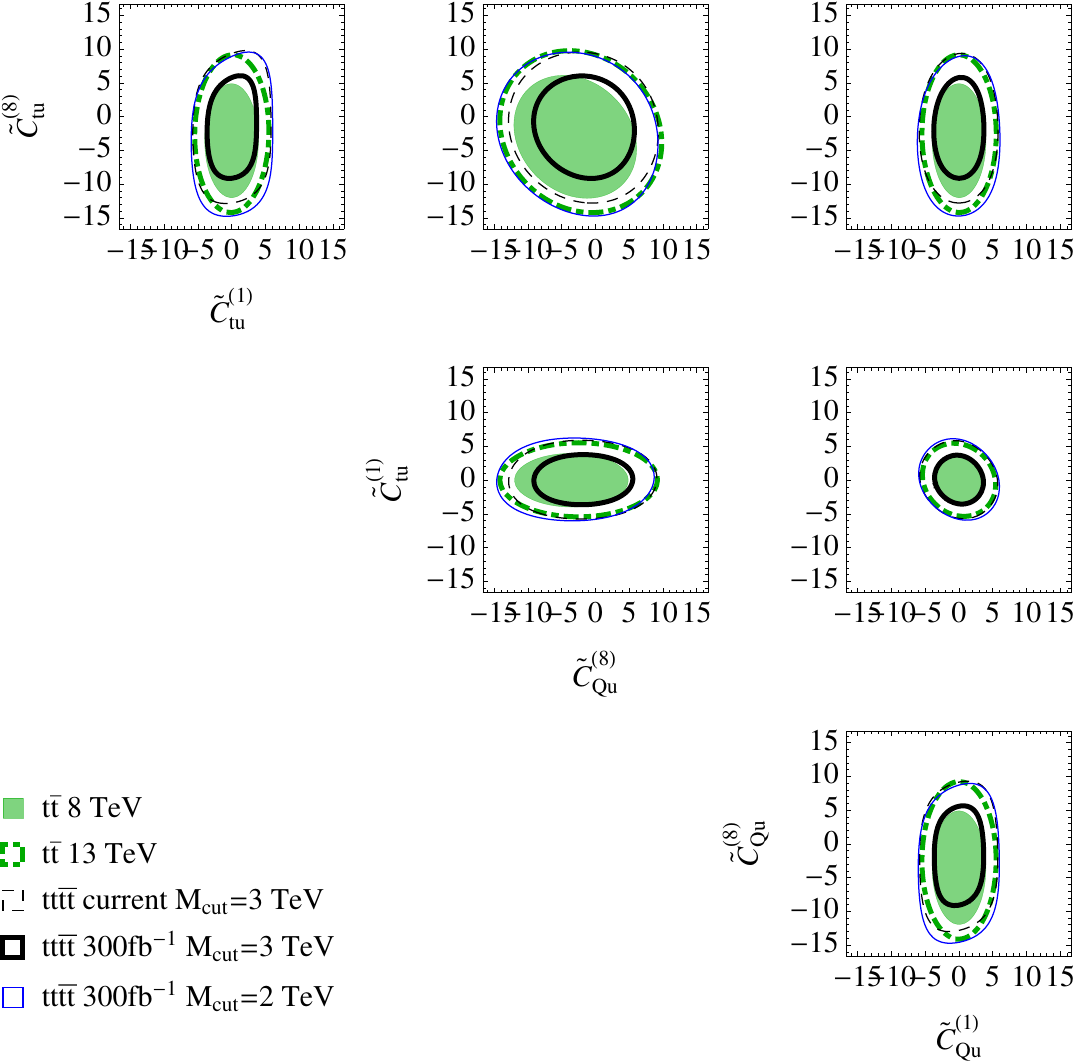}
\figcaption{ \label{fig:limitsup}
Constraints from four-top cross section and individual $t\bar t$ cross section
measurements, on the operator coefficients in the first category ($\tilde
C^{(8)}_{tu}$, $\tilde C^{(1)}_{tu}$, $\tilde C^{(8)}_{Qu}$, $\tilde
C^{(1)}_{Qu}$), assuming two coefficients to be nonzero at a time.
}
\end{center}

It is important to point out that the $t\bar t$ measurement is limited by
systematic errors, and further improvements with higher luminosity is
difficult.  On the other hand, there is still a lot of room for the four-top
search/measurement to be improved in the future.  Our results indicate that, in
the near future, the four-top process could even
take place and provide more crucial information on
$qqtt$ operators.  One should keep in mind that this requires a relatively
large cut, $M_{cut}\gtrsim2\sim3$ TeV, to be applied in the analysis, so only
the BSM models that live above $\sim3$ TeV are subject to these new
constraints.  This is however not a drawback from the SMEFT point of view,
because new states below this energy scale are likely to be excluded by
explicit resonance searches \cite{Patrignani:2016xqp}.

\section{Global fit}\label{sec:global}

In the previous section we have shown that the future four-top measurement can
have a better sensitivity to four-fermion operators compared with individual
$t\bar t$ cross section measurements.  However, the $t\bar t$ production is
such an important process that has been measured extensively at many different
energies (Tevatron, LHC 7, 8, 13 TeV) and in many different ways (cross section,
asymmetries, distributions, etc.)  A global fit of all available measurements
thus provides so far the best available limits.  In this section we will
investigate whether the four-top process can add useful information on top of
the global top measurement program.

The recent global fit performed by the authors of
Refs.~\cite{Buckley:2015nca,Buckley:2015lku} has included the four-fermion
operators.  The fit is based on four linear combinations of their coefficients,
called $C_{u,d}^{1,2}$, which are the only independent degrees of freedom at
the dim-6 interference level \cite{Zhang:2010dr}. However, the theoretical set
up in this work is different, in that we expect that the dim-6 squared terms
can dominate given the current bounds.  This has been confirmed by, for
example, a fit for four-fermion operators in Ref.~\cite{Rosello:2015sck}, where
this domination has been interpreted as the SMEFT being invalid.  However, as
explained in Section~{\ref{sec:sensi}}, in this work we distinguish the
$CE^2/\Lambda^2$ expansion from the true EFT expansion $E^2/\Lambda_{NP}^2$,
and in any case we expect higher powers of $CE^2/\Lambda^2$ to dominate in
four-top production and to enhance its sensitivity to BSM.  The actual EFT
validity is then guaranteed by the assumption $\Lambda_{NP}>M_{cut}\sim$ a few
TeV, which is more than enough for most $t\bar t$ measurements.

In light of the above considerations, the fit we will perform is different than
the previous ones, in that the dim-6 squared terms as well as the interference
between two dim-6 operators will be fully incorporated.  We will have to abandon
the $C_{u,d}^{1,2}$ language, as this simplification breaks down at the
dim-6 squared level.  Furthermore, the color-singlet operators cannot be neglected
due to vanishing interference term.  The fit will then include 14 independent
operators.  A complete analysis including every existing measurements is
certainly beyond the scope of this work.  Given that the goal is to evaluate
the relative constraining powers of $t\bar t$ and four-top production,
we will follow the approach in Ref.~\cite{Rosello:2015sck}, where the most relevant
measurements on cross sections and asymmetries have been included. 
The most recent LHC 13 TeV cross section measurements will be added as well.
Furthermore, unlike Ref.~\cite{Rosello:2015sck}, we shall also consider the
differential $m_{tt}$ distribution measurement to constrain the possible shape
change from four-fermion operators
\cite{Degrande:2010kt}.  In Table~\ref{tab:data} we list the measurements that
will be used in our fit, together with the corresponding SM predictions. 
We believe these observables represent the most sensitive ones to $qqtt$ operators
that have been measured so far.

For simplicity, we only consider the SM prediction uncertainties as theoretical
uncertainties.  We add the theoretical and experimental errors in quadrature, and when
the errors are not symmetric, we take the larger one for both sides.  We
further assume all uncertainties are not correlated, except for the theory ones
that come from the same prediction.  The fit for the $m_{tt}$
distribution should be considered at most as a ``toy fit'' given that
correlations between different bins are not available from the experimental
report.  We have dropped the last bin due to the normalization constraint.  A
$K$-factor rescaling will not improve the normalized distribution.  For this
reason, we use LO prediction and consider various theory errors. The scale
uncertainty is from variation of $\mu_R$ and $\mu_F$ by a factor of 2.  The PDF
uncertainty is taken from the envelope of three PDFs, including the NNPDF
\cite{Ball:2014uwa}, MMHT \cite{Harland-Lang:2014zoa}, and CT14
\cite{Dulat:2015mca} PDF sets with their own uncertainty bands.  We have
checked that the differences between LO and NLO predictions are within these
errors.
\end{multicols}

\vspace*{3.5cm}

\begin{center}
\tabcaption{ \label{tab:data}  Measurements used in the global fit, with
corresponding theory predictions and uncertainties.}
\footnotesize
\begin{tabular*}{170mm}{@{\extracolsep{\fill}}lcc}
\toprule
  & SM prediction & Measurement \\
\hline
Cross section, Tevatron 1.96 TeV, CDF+D0 
& 7.35$^{+0.26}_{-0.33}$ pb  \cite{Czakon:2013goa} & 7.60$\pm$0.41 pb \cite{Aaltonen:2013wca}\\
\\
Cross section, LHC 8 TeV, ATLAS+CMS
& 252.9$^{+13.3}_{-14.5}$ pb  \cite{Czakon:2013goa} & 241.5$\pm$8.5 pb \cite{CMS:2014gta}\\
\\
Cross section, LHC 13 TeV, CMS
& 832$^{+40}_{-45}$ pb  \cite{Czakon:2011xx} & 888$^{+33}_{-34}$ pb \cite{Sirunyan:2017uhy}\\
\\
Cross section, LHC 13 TeV, ATLAS
& 832$^{+40}_{-45}$ pb  \cite{Czakon:2011xx} & 818$^{+36}_{-36}$ pb \cite{Aaboud:2016pbd}\\
\\
$A_{FB}$, Tevatron 1.96 TeV, CDF+D0 
& 0.095 $\pm$ 0.007  \cite{Czakon:2014xsa} & 0.128 $\pm$ 0.025 \cite{CDF:2016pir}\\
\\
$A_{C}$, LHC 8 TeV, ATLAS 
& 0.0111 $\pm$ 0.0004 \cite{Bernreuther:2012sx} & 0.009 $\pm$ 0.005 \cite{Aad:2015noh}\\
\\
$A_{C}$, LHC 8 TeV, CMS
& 0.0111 $\pm$ 0.0004 \cite{Bernreuther:2012sx} & 0.0033 $\pm$ 0.0042
\cite{Khachatryan:2015mna}\\\hline
\\
$m_{tt}$ distribution, LHC 8 TeV, ATLAS & {\sc MadGraph5\_aMC@NLO+PYTHIA6} \cite{Sjostrand:2000wi} & Ref.~\cite{Aaboud:2016iot}
\\
\bottomrule
\end{tabular*}%
\end{center}
\begin{multicols}{2}

A $\chi^2$ is constructed based on the information in Table~\ref{tab:data},
to derive the 95\% CL limits on operator coefficients.
These limits are compared with the projection of four-top measurement
at 300 fb$^{-1}$.  Some results are shown in Figure~\ref{fig:ttvstttt}, with
two operators turned on at a time.  Our observations are the following:
\begin{itemize}
\item In general, with a $3\sim 4$ TeV cut, constraints from four-top cross
	section are very similar to those from $m_{tt}$ measurement (black
	dashed vs blue and red).  In rare cases they are complementary.
\item In most cases, combining the $t\bar t$ inclusive measurements, i.e.~the
cross sections and the asymmetries, provides the most constraining
limits, as expected.  This is illustrated by the three plots
in the first row in Figure~\ref{fig:ttvstttt}.
Results from $m_{tt}$ differential measurements and four-top cross sections
provide slightly weaker bounds.  The $t\bar t$ global fit, including cross sections,
asymmetries, and $m_{tt}$, is indicated by the green shaded area.

\item
Exceptions are the directions that are not effectively constrained by asymmetry
measurements.  In this case both $m_{tt}$ differential measurements and the
four-top cross section provide better limits.  These cases are illustrated in
the second row in Figure~\ref{fig:ttvstttt}.  The diagonal directions roughly
correspond to flat directions between the LLLL/RRRR operators and the LLRR/RRLL
operators, whose contributions to $A_{FB}$ and $A_C$ have the opposite sign.
The asymmetry measurements thus do not provide useful information in these
directions.  When two operators are turned on, there can be four such
directions, as the dominant contributions are coming from dim-6 squared terms.
This can be seen in Figure~\ref{fig:ttvstttt}, second row.  Clearly, in these
cases both the $m_{tt}$ and the four-top measurements help to further improve our
reach in SM deviations.
\end{itemize}

Given that the four-top measurement provides almost the same information as the
$m_{tt}$ differential cross section, we do not expect the four-top to give
better constraints than a global fit on the $qqtt$ operators.  It is however a
valuable add to the precision top physics at the LHC, given that the $m_{tt}$
distribution is already one of the most sensitive observables to
four-fermion operators.  In particular, in directions that are not sensitive
to asymmetry measurements, information from four-top process is useful.
For this reason we expect that marginalized constraints from a $t\bar t$ global
fit and those from the four-top process are comparable.  However, to really
confirm this point in a model-independent way, we need to take into account the
four-top operators given in Eqs.~(\ref{eq:4t1})-(\ref{eq:4t2}), to derive the
fully marginalized constraints from the four-top process.  Naively, one would
expect that further marginalizing over the additional $tttt$ operators would
make the constraints on $qqtt$ operators weaker.  We will show that, while this
is indeed the case, the effect is not large enough to qualitatively change our
conclusion.

\end{multicols}
\newpage

\bigskip
\ruleup

	\begin{center}
		\includegraphics[width=.9\linewidth]{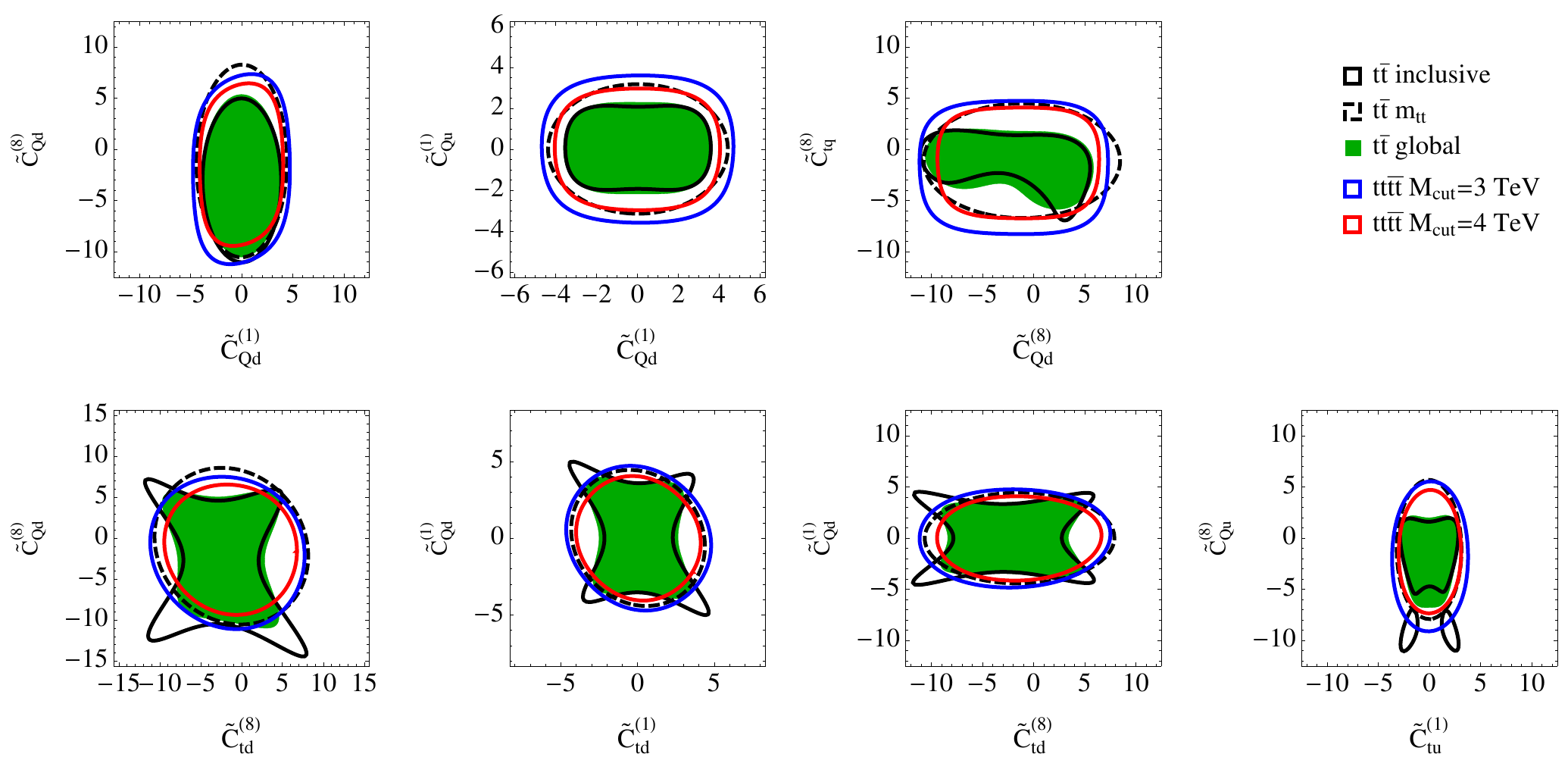}
	\figcaption{\label{fig:ttvstttt}
Selected results for $t\bar t$ global fit, compared with 
projected constraints from the four-top production at high luminosity,
at 95\% CL.  Black solid and dashed contours represent constraints
from $t\bar t$ inclusive measurements (i.e.~cross sections and asymmetries) and
$m_{tt}$ differential measurement respectively.  The green shaded area
is the combined result.  Constraints from $tt\bar t\bar t$ with $M_{cut}=3$, 4
TeV are given by the blue and red curves separately.
	}
	\end{center}
\ruledown
\bigskip
\begin{multicols}{2}

\begin{center}
		\includegraphics[width=\linewidth]{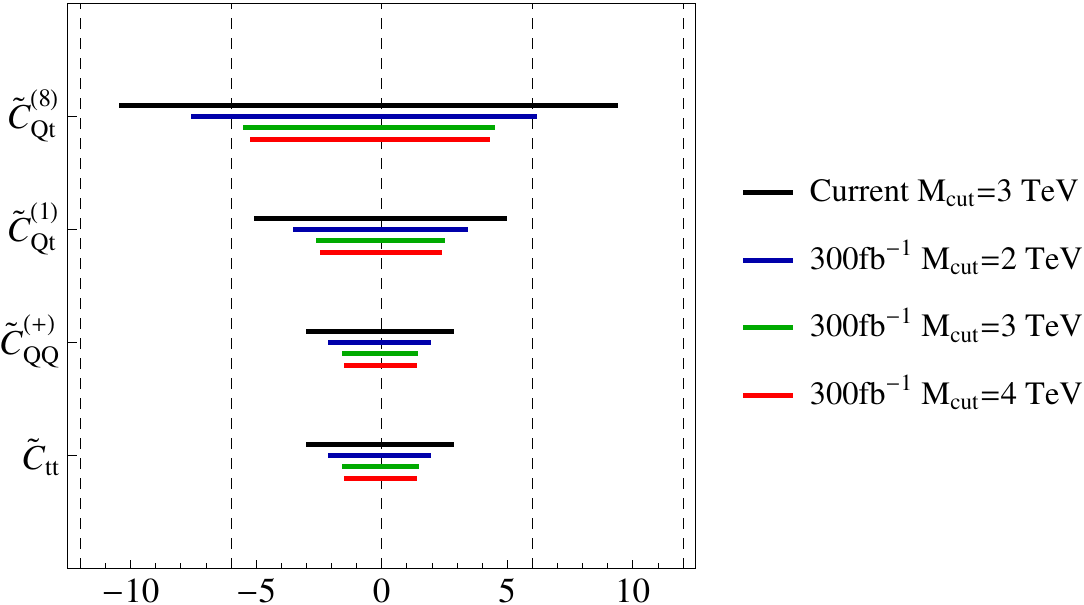}
		\figcaption{\label{fig:4tlimit}
		Marginalized constraints on four-top operators,
		using the current bound as well as the projection
		for 300 fb$^{-1}$, for several $M_{cut}$ values.}
\end{center}

For illustration, in Figure~\ref{fig:4tlimit} we present the constraints on
four-top operators.  These are marginalized over other four-top operators,
but not the $qqtt$ operators.  The current constraints are derived using
$M_{cut}=3$ TeV, while the projected ones for 300 fb$^{-1}$ are given with
$M_{cut}=2$, 3, 4 TeV, respectively.  The constraints are more conservative
than those directly extracted from a tailored experimental analysis,
e.g.~Ref.~\cite{ATLAS:2016btu}.  This is expected because we assumed SM signal
shape and only use the cross section below $M_{cut}$.  Also note that even for
the most constraining limits, the dim-6 squared contribution already dominates
over the interference.  For this reason, including these operators in our
analysis should not significantly affect the constraints on the other 14 $qqtt$
operators.

In Figure~\ref{fig:con} we present the most important results of this work:
a comparison of fixed (i.e.~one operator at a time) and fully marginalized
(i.e.~all other operators floated) constraints for all $qqtt$ operators, from
the four-top measurement and from the $t\bar t$ measurements.  The $t\bar t$
constraints are from our global fit, including cross sections, asymmetries,
and $m_{tt}$ distribution, while the four-top constraints are from
the 300 fb$^{-1}$ projection, $\mu<1.87$, with different $M_{cut}$ values
applied.  Perturbativity in the EFT requires Eq.~(\ref{eq:pert}) to hold.  This
leads to an upper bound of $|\tilde C|<39$, 18, and 9.9 respectively for
$M_{cut}=2$, 3, 4 TeV.  The latter two are shown in Figure~\ref{fig:con} by the
vertical dotted lines.

\begin{center}
	\vspace{.5cm}
		\includegraphics[width=\linewidth]{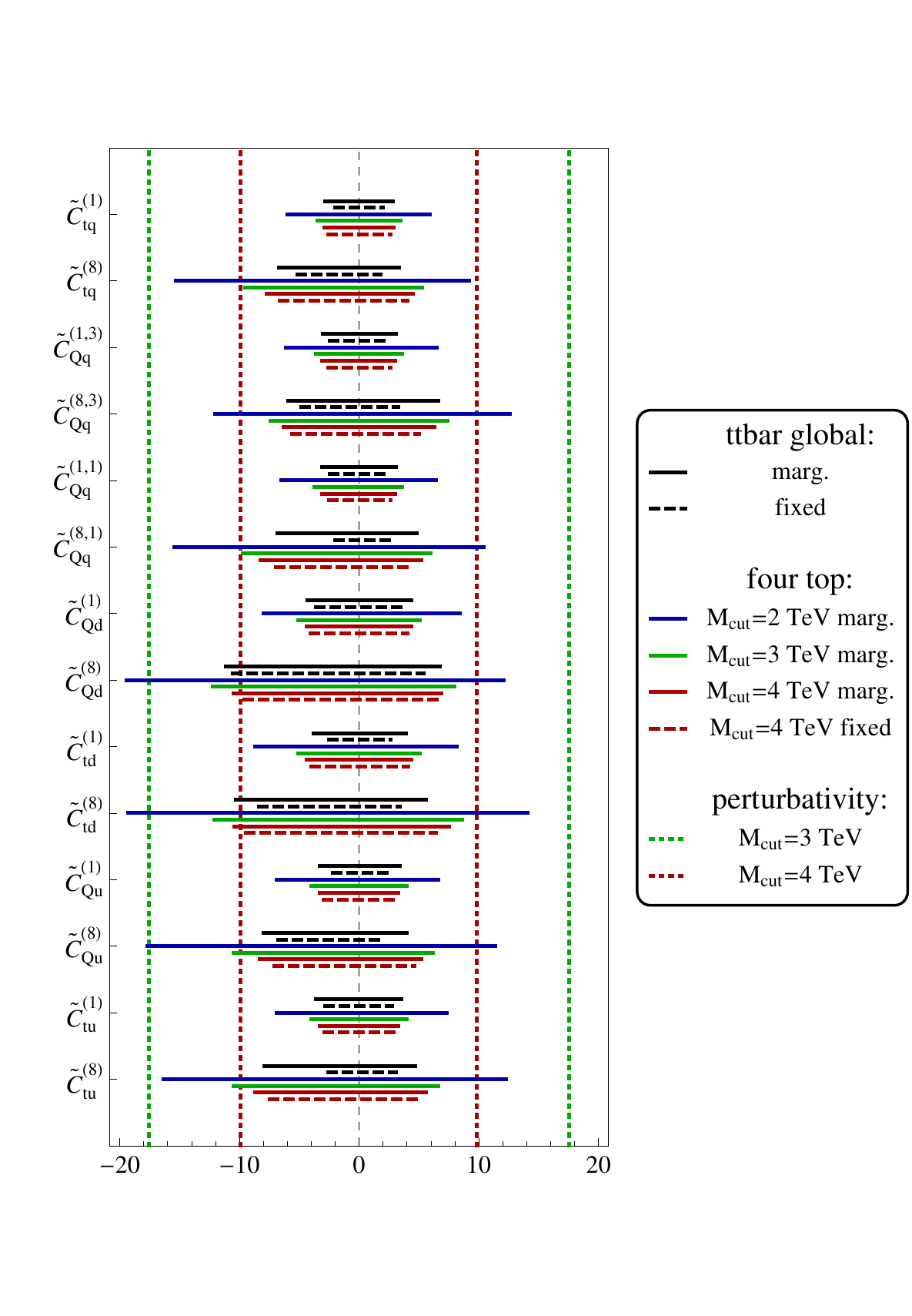}
		\figcaption{\label{fig:con}
Fixed (i.e.~one operator at a time) and fully marginalized
(i.e.~all other operators floated) constraints for all $qqtt$ operators, from
four-top measurement and from $t\bar t$ measurements, at 95\% CL.  The $t\bar t$
constraints are from our global fit, while the four-top constraints are form
the 300 fb$^{-1}$ projection. Different $M_{cut}$ values are applied.
Perturbativity bounds are derived from Eq.~(\ref{eq:pert}).  }
\end{center}

We can see that while the $M_{cut}=2$ TeV results are worse, the $M_{cut}=3$, 4
TeV marginalized constraints are in general as good as those from the $t\bar t$
global fit.  This agrees with our previous expectation: the $t\bar t$ global
analysis gives better individual constraints, thanks to the asymmetry
measurements, while the four-top production is very helpful in directions that
are not sensitive to these measurements.  Marginalizing over additional
four-top operators does not significantly change our results.  In fact, for the
four-top process the difference between individual limits and fully marginalized
ones is in general not very large (see the difference between the red solid 
and the red dashed lines), which implies that the cross section is dominated by
the $(CE^2/\Lambda^2)^4$ terms, while the interference between different
operators or with the SM is small.  We want to emphasize that the four-top
constraints obtained in this work are in general conservative, and in practice
better results can be expected from a tailored experimental analysis.  For
example, in the case of four-$t_R$ operator $\mathcal{O}_{tt}$, by assuming the
spectrum via the EFT model, the constraints can be enhanced by a factor of
$\sim2$ compared with assuming the SM signal shape \cite{ATLAS:2016btu}.  One
could imagine that a similar factor applies also to $qqtt$ operators, and in
this case the four-top cross section could even be more constraining than a
$t\bar t$ global fit.  In addition, in the future combining
searches/measurements in different channels could further improve the reach.  

The perturbativity in general is not a problem for $M_{cut}=2$, 3 TeV, while
for 4 TeV some of the marginalized constraints start to approach or even go
beyond the perturbative limit.  Unitarity gives further constraints.  Following
Ref.~\cite{Degrande:2012wf}, we find that the following constraints
\begin{flalign}
	&C^{(1)}_i\frac{E^2}{\Lambda^2}\lesssim 4\sqrt{6}\pi\,,
	\\
	&C^{(8)}_i\frac{E^2}{\Lambda^2}\lesssim 24\sqrt{2}\pi\,,
\end{flalign}
apply to color singlet and octet operators respectively.  These values
imply that the limits obtained with $M_{cut}=2\sim3$ TeV are more or less safe,
while the improvements due to including events from $3\sim4$ TeV are not reliable,
even though these improvements are quite small already.
It should be noted that the unitarity problem with $M_{cut}\approx 4$ TeV
is at most temporary, as the experimental precision will continue to improve.
In fact, considering the possible improvements discussed in the previous
paragraph, it is likely that going to 4 TeV is safe already with the assumed
luminosity.

In any case, we conclude that in the near future, compared with a $t\bar t$
global fit, the four-top production can provide competitive constraints, due to
its enhanced sensitivity to four-fermion $qqtt$ operators.  Including this
process in the global top-fitting program will definitely improve our reach in
SM deviations in the top sector.  Compared with $t\bar t$ inclusive
measurements which are dominated by systematic errors, the high-mass $t\bar t$
production and the four-top production have more room to improve. In the long
term, they should become the crucial ones that determine our final reach at the
LHC.

Before concluding, we remind the reader that the cost of such an enhanced
sensitivity is a relatively large value of $M_{cut}$, which implies that the
results are applicable only to BSM scenarios above this scale.  In the long
term, however, we believe that in any case new states below this energy scale
are likely to be excluded by explicit resonance searches.  One should also keep
in mind that when these results are interpreted with explicit BSM models, it
typically implies that a large BSM coupling is allowed, and so one should
always check the sizes of higher dimensional operators, to make sure that the
truncation of the SMEFT at dim-6 is valid.  

\section{Conclusion}\label{sec:conclusion}

Precision measurements are not just for precision itself.  The ultimate goal is
the higher reach in testing new physics and the ability to exclude deviations
from the SM, and therefore sensitivity to SM deviations is crucial.  An
observable with an enhanced sensitivity to SM deviations, even poorly measured,
may have a chance to play an important role.  We have demonstrated
this last point, using the four-top production process in the top EFT context.
As a benchmark to assess its sensitivity, we use the top-pair production
measurements for comparison.  We have found that, as far as the dim-6
four-fermion operators are concerned, the current upper bound on the four-top
signal strength at the $\mathcal{O}(10)$ level is already as powerful as
the $t\bar t$ cross section measurements which have percentage level precision.
Furthermore, using the projected bounds for 300 fb$^{-1}$ at 13 TeV, the
four-top measurement can even compete with a global fit using $t\bar t$
measurements, including the $m_{tt}$ differential distributions.  This
comparison is remarkable as the four-top cross section was never considered as
a precision measurement like the $t\bar t$ process.

The origin of the enhanced sensitivity of four-top cross section comes from
the fact the four-fermion operators can be inserted up to four times in the
squared amplitude, each time with a factor $CE^2/\Lambda^2$ enhancing their
contribution to the cross section. This factor can be larger than one given the
current limits on $C/\Lambda^2$, and the typical center-of-mass energy of
the four top quarks produced.  We have shown that the validity of the EFT, or in
other words the validity of expansion in higher dimensional operators, can be
controlled by $E^2<M_{cut}^2<\Lambda_{NP}^2$, without spoiling the enhancement
effect for $M_{cut}\sim$ a few TeV, and that the EFT perturbativity
$CM_{cut}^2/\Lambda^2<(4\pi)^2$ is also satisfied in general.  The four-top
measurement can thus provide useful bounds for underlying BSM models that live
at a scale $\gtrsim$ a few TeV, and is therefore a valuable add to the
precision top physics at the LHC, in particular, given that there is still a
lot of room for this process to improve in the future.  On the other hand, for
BSM scenarios below a few TeV, these results may not apply, but in any case one
expects that there the explicit resonant searches provide better exclusion.

For comparison purpose we have performed a global fit for the most relevant
$t\bar t$ measurements, including a differential measurement on $m_{tt}$, to
which the four-fermion operators are sensitive.  Unlike previous studies, our
fit is done including all dim-6 squared terms as well as interference effects
between all 14 dim-6 operators.  We have also included the four-top operators
in our analysis, and have demonstrated that marginalizing over these operators
do not qualitatively change our conclusion. Compared with our $t\bar t$ global
fit, the four-top process gives comparable limits on all operator coefficients.
One should however keep in mind that these limits are still relatively
conservative, given that the upper bound on the total cross section assumes SM
signal shape, and is used regardless of the value of the $M_{cut}$.  We expect
that future experimental analyses following the SMEFT strategy will further
improve the sensitivity of this process to SM deviations.

Finally, we would like to point out that potentially other processes can have a
similar enhanced sensitivity, provided that the following conditions are
satisfied: 1) there are multiple heavy particles in the final state, so that
the process is naturally related with a large energy scale; 2) multiple
insertion of dim-6 operators are allowed, and thus potentially leading to more
powers of $CE^2/\Lambda^2$ enhancing the EFT contribution; and 3) the contribution
of dim-6 operators goes like $E^2/\Lambda^2$, i.e.~not suppressed by any mass
or Higgs vev factors.  Of course, the validity of EFT has to be checked
carefully as one starts to approach the boundary of its applicability.
Still, we hope that this study could inspire new ideas about using observables
that are not so precisely measured, to further push the frontier of precision
measurements in the EFT context.

\section*{Acknowledgements}

CZ thanks Gauthier Durieux and Fabio Maltoni for their invaluable advice.

\appendix
\begin{small}
\section*{Appendix A: Operator basis}

\newcommand*{\tmp}[4]{%
	{#4%
		\ifx\empty#3\empty\ifx\empty#1\empty\else^{(#1)}\fi
		\else\ifx\empty#1\empty^{(#3)}\else^{(#1)(#3)}\fi\fi%
	\ifx\empty#2\empty\else_{#2}\fi}%
}
\newcommand*{\cc }[3]{\tmp{#1}{#2}{#3}{C}}
\newcommand*{\ccc}[3]{\tmp{#1}{#2}{#3}{c}}
Here we present the relations between the coefficients of 
our four-fermion operators and those of the basis
operators in the so-called Warsaw basis, in Ref.~\cite{Grzadkowski:2010es}.

\begin{subequations}
\renewcommand{\theequation}{A\arabic{equation}}
\begin{flalign}
	&\mbox{$qqtt$ operator coefficients:}\nonumber
	\\
	&\cc{1,8}{Qq}{}\equiv
	\cc{1}{qq}{i33i}+3\cc{3}{qq}{i33i},
	\label{eq:octet_1}
	\\
	&\cc{3,8}{Qq}{}\equiv
	\cc{1}{qq}{i33i}-\cc{3}{qq}{i33i},
	\label{eq:octet_2}
	\\
	&\cc{1,1}{Qq}{}\equiv
		\cc{1}{qq}{ii33}
		+\frac{1}{6}\cc{1}{qq}{i33i}
		+\frac{1}{2}\cc{3}{qq}{i33i},
	\label{eq:singlet_1}
	\\
	&\cc{3,1}{Qq}{}\equiv
		\cc{3}{qq}{ii33}
		+\frac{1}{6}\left(\cc{1}{qq}{i33i}-\cc{3}{qq}{i33i}\right),
	\label{eq:singlet_2}
	\\
	&\cc{8}{tu}{} \equiv 2 \cc{}{uu}{i33i},
\\
	&\cc{8}{td}{} \equiv \cc{8}{ud}{33ii},
	\\
	&\cc{1}{tu}{} \equiv \cc{}{uu}{ii33} +\frac{1}{3} \cc{}{uu}{i33i},
\\
	&\cc{1}{td}{} \equiv \cc{1}{ud}{33ii},
	\\
	&\cc{8}{tq}{}\equiv\cc{8}{qu}{ii33},
	\\
	&\cc{8}{Qu}{}\equiv\cc{8}{qu}{33ii},
	\\
	&\cc{8}{Qd}{}\equiv\cc{8}{qd}{33ii},
	\\
	&\cc{1}{tq}{}\equiv\cc{1}{qu}{ii33},
	\\
	&\cc{1}{Qu}{}\equiv\cc{1}{qu}{33ii},
	\\
	&\cc{1}{Qd}{}\equiv\cc{1}{qd}{33ii};
	\\
	&\mbox{$tttt$ operator coefficients:}\nonumber
	\\
	&\cc{+}{QQ}{}\equiv\cc{1}{qq}{3333}+\cc{3}{qq}{3333},
	\\
	&\cc{1}{tt}{}\equiv\cc{}{uu}{3333},
	\\
	&\cc{1}{Qt}{}\equiv\cc{1}{qu}{3333},
	\\
	&\cc{8}{Qt}{}\equiv\cc{8}{qu}{3333},
\end{flalign}
\end{subequations}
where on the l.h.s are the coefficients of the operators used in this work,
while on the r.h.s are the coefficients of the Warsaw operators. $i=1,2$
is a flavor index.

\section*{Appendix B: More results}

We present constraints from four-top production and $t\bar t$ cross section
measurements, similar to Figure~\ref{fig:limitsup}, but for the operators in
the 2nd and the 3rd categories, i.e.~in Eqs.~(\ref{eq:c2}) and (\ref{eq:c3}).
They are displayed in Figure~\ref{fig:limitsd} and Figure~\ref{fig:limitsq},
respectively.

	\begin{center}
		\includegraphics[width=\linewidth]{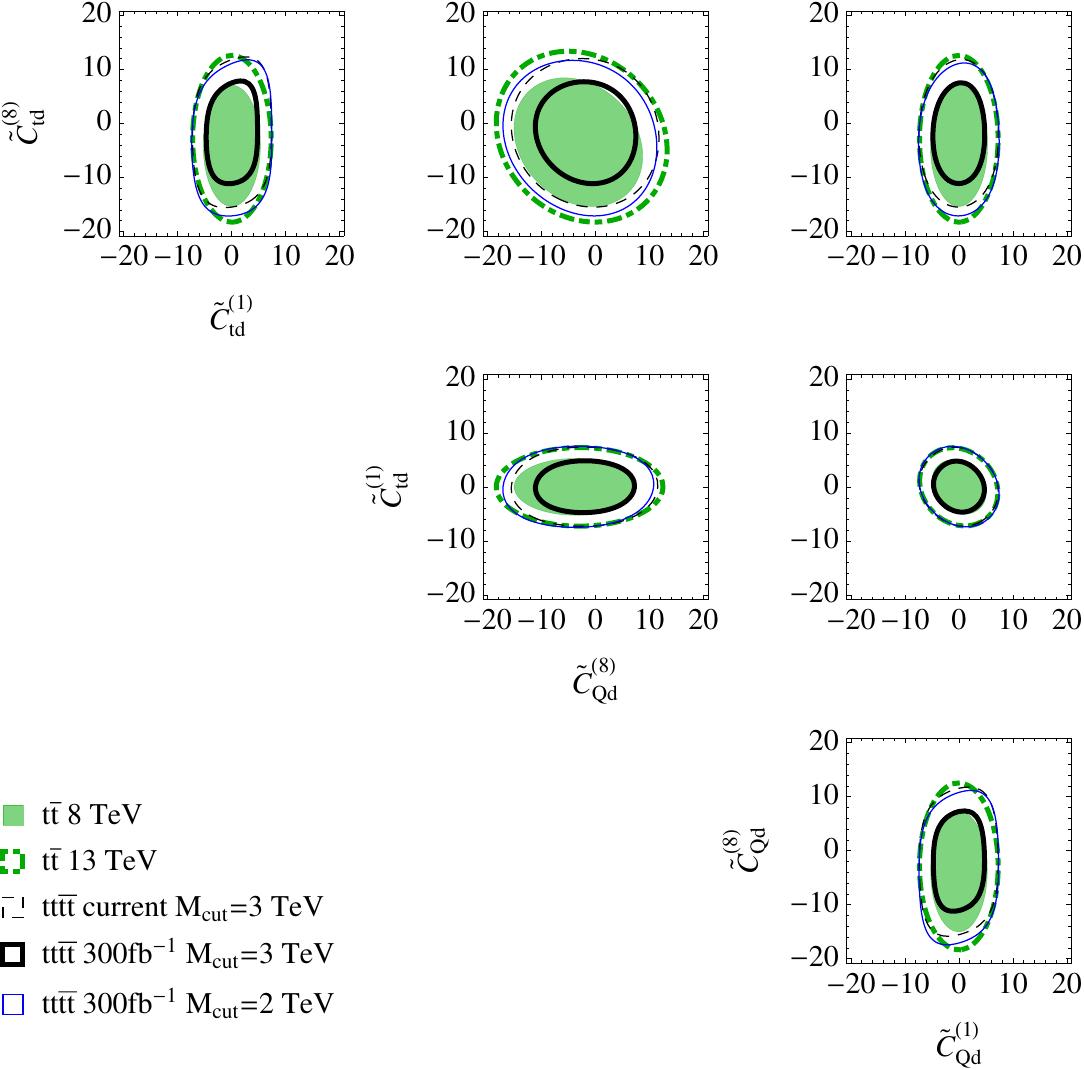}
		\\*
	\figcaption{ \label{fig:limitsd}
	Constraints from four-top cross section and individual $t\bar t$ cross
	section measurements, on the operator coefficients in the second
	category ($\tilde C^{(8)}_{td}$, $\tilde C^{(1)}_{td}$, $\tilde
	C^{(8)}_{Qd}$, $\tilde C^{(1)}_{Qd}$), assuming two coefficients
	to be nonzero at a time.
	}
	\end{center}
	\vspace*{9cm}
\end{small}
\end{multicols}
\bigskip
\ruleup
	\begin{center}
		\includegraphics[width=.8\linewidth]{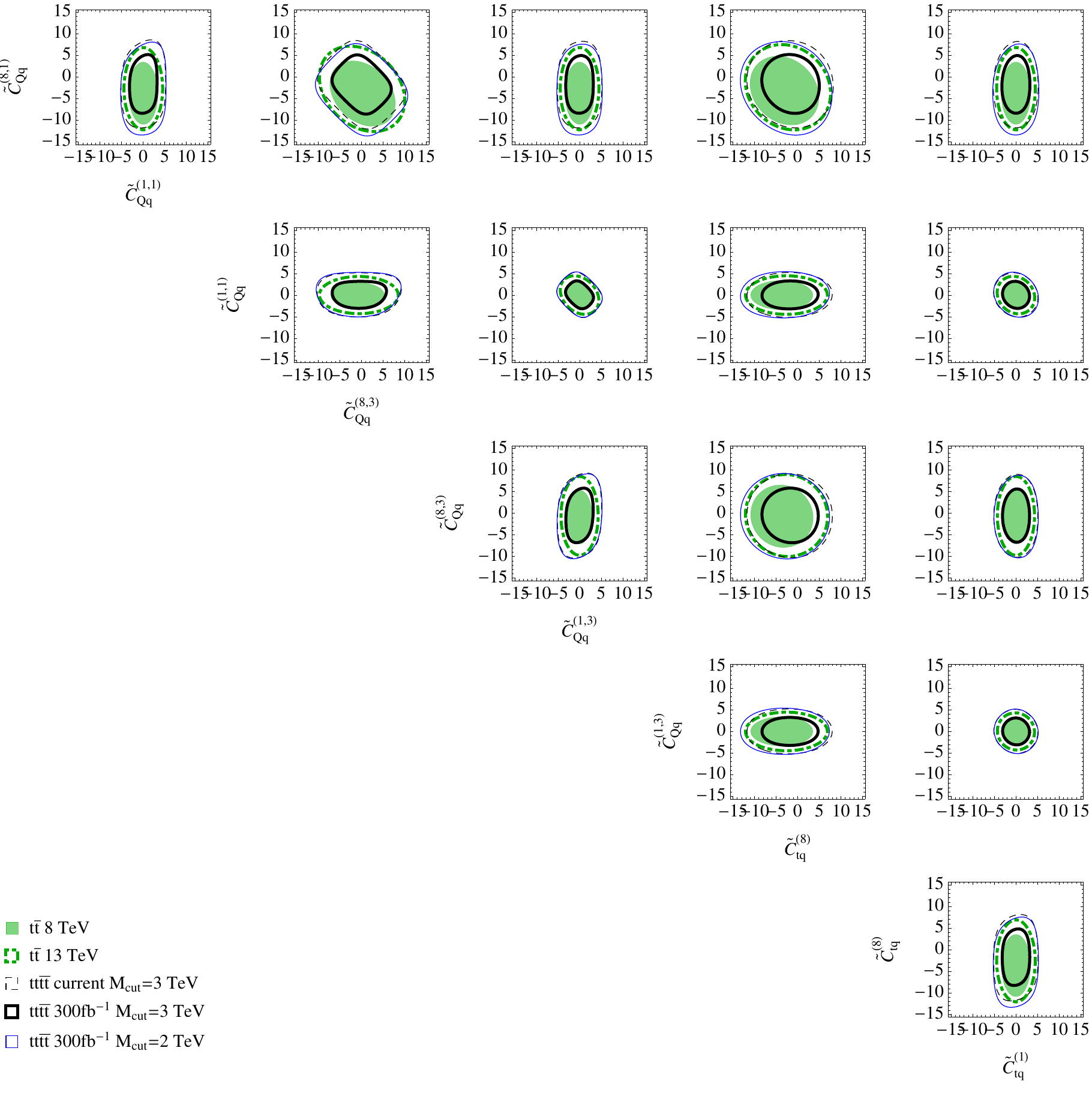}
	\figcaption{ \label{fig:limitsq}
	Constraints from four-top cross section and individual $t\bar t$ cross
	section measurements, on the operator coefficients in the third
	category ($\tilde C^{(8,1)}_{Qq}$, $\tilde C^{(1,1)}_{Qq}$,
	$\tilde C^{(8,3)}_{Qq}$, $\tilde C^{(1,3)}_{Qq}$,
	$\tilde C^{(8)}_{tq}$, $\tilde C^{(1)}_{tq}$), assuming two coefficients
	to be nonzero at a time.
	}
	\end{center}
\ruledown
\bigskip
\begin{multicols}{2}

\begin{small}

\end{small}

\end{multicols}
\end{document}